\documentclass{iopart}

\usepackage{graphicx}
\usepackage{bbm}
\usepackage{enumerate}

\usepackage{epstopdf}
\newcommand{\be}{\begin{equation}}
\newcommand{\ee}{\end{equation}}

\newcommand{\eev}{~\mathrm{EeV}}

\newcommand{\bea}{\begin{eqnarray}}
\newcommand{\eea}{\end{eqnarray}}

\begin{document}

\title{Positron decrement in AMS-02 compared to Pamela, due to solar cycle}

\author{Jonathan P. Roberts}
\address{Center for Cosmology and Particle Physics, Department of Physics\\
New York University, NY, NY 10003, USA\\}

\begin{abstract}
We present a numerical simulation of cosmic ray transport in the heliosphere and show that AMS-02 should observe a decrease in the positron fraction above 40GeV compared to the fraction observed by PAMELA.
\end{abstract}

\section{Introduction}
In 2008 the PAMELA satellite reported an observation of many more positron cosmic rays between 10 and 100GeV than predicted from models of cosmic ray transport \cite{Adriani:2008zr}. In \cite{JR PAMELA} we proposed that the observed positron excess was due to the configuration of the heliospheric magnetic field. During periods of solar minimum cosmic rays of one charge can penetrate the center of the solar system whereas cosmic rays of the opposite charge are deflected. When the sun enters a period of solar maximum, the magnitude of the excess will decrease and eventually change sign, creating an excess of the opposite charge. A measurement by AMS-02 of a decrease in the positron excess correlated to an increase in the tilt angle of the heliospheric current sheet would be a smoking gun of a heliospheric origin of the positron excess.

\section{The current sheet and particle transport}

\begin{figure}
\begin{center}
\scalebox{0.8}{\includegraphics{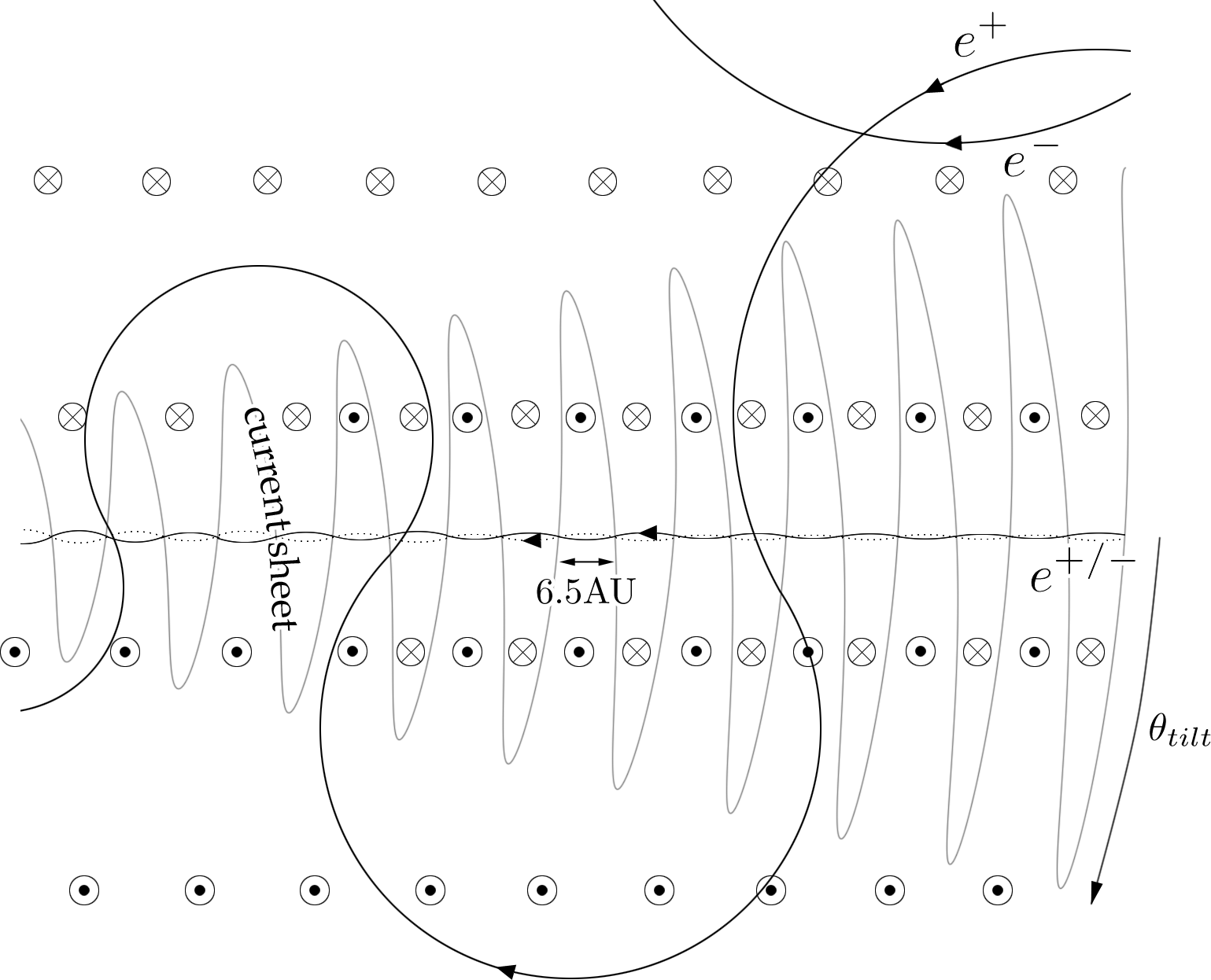}}
\end{center}
\caption{A schematic of the heliospheric magnetic field. The rapidly oscillating grey line indicates the current sheet. The latitudinal extent of the current sheet is specified by the angle $\theta_{tilt}$. The center of the solar system is to the left. The magnetic field orientation matches the orientation during the previous solar minimum. Lines with arrows indicate possible trajectories of electrons and positrons. \label{MagneticField}}
\end{figure}

The current sheet is the plane that divides magnetic fields of opposite orientations in the heliosphere. If the solar magnetic poles were perfectly aligned with the axis of rotation, the current sheet would be a flat plane perpendicular to the rotation axis. As the sun's magnetic poles are not aligned with the rotation axis the current sheet oscillates, creating a `ballerina skirt' shape. The latitudinal extent of the oscillations is determined by the angle between the magnetic pole and the axis of rotation, $\theta_{tilt}$. A schematic of the magnetic field is shown in Fig.~\ref{MagneticField}. The large scale magnetic field is spiral and approximates a toroidal field at larger radii ($>10AU$), with a strength that falls off as $1/r$ \cite{UlyssesBook}.

Levy \cite{Levy1,Levy2} showed that a flat current sheet could produce a charge asymmetry in the flux of cosmic rays with energies of a few tens of GeV. We showed that an equivalent charge asymmetry exists in a magnetic field configuration with a fluctuating current sheet \cite{JR PAMELA}. The origin of the charge asymmetry is apparent from the particle paths illustrated in Fig.~\ref{MagneticField}. Every time a particle crosses the current sheet the direction of curvature reverses - directing positrons back towards the ecliptic plane, but directing electrons away. For particles that remain close to the ecliptic plane, there is a charge symmetric route to the center of the heliosphere. At high latitudes, only positrons can cross the ecliptic multiple times and penetrate the center of the heliosphere, whereas electrons are deflected away from the current sheet and are trapped in the toroidal fields. The fraction of positrons that have paths that cross the current sheet is proportional to the Larmor radius, resulting in a positron fraction that increases with energy.

\begin{figure}
\begin{center}
\scalebox{0.8}{\includegraphics{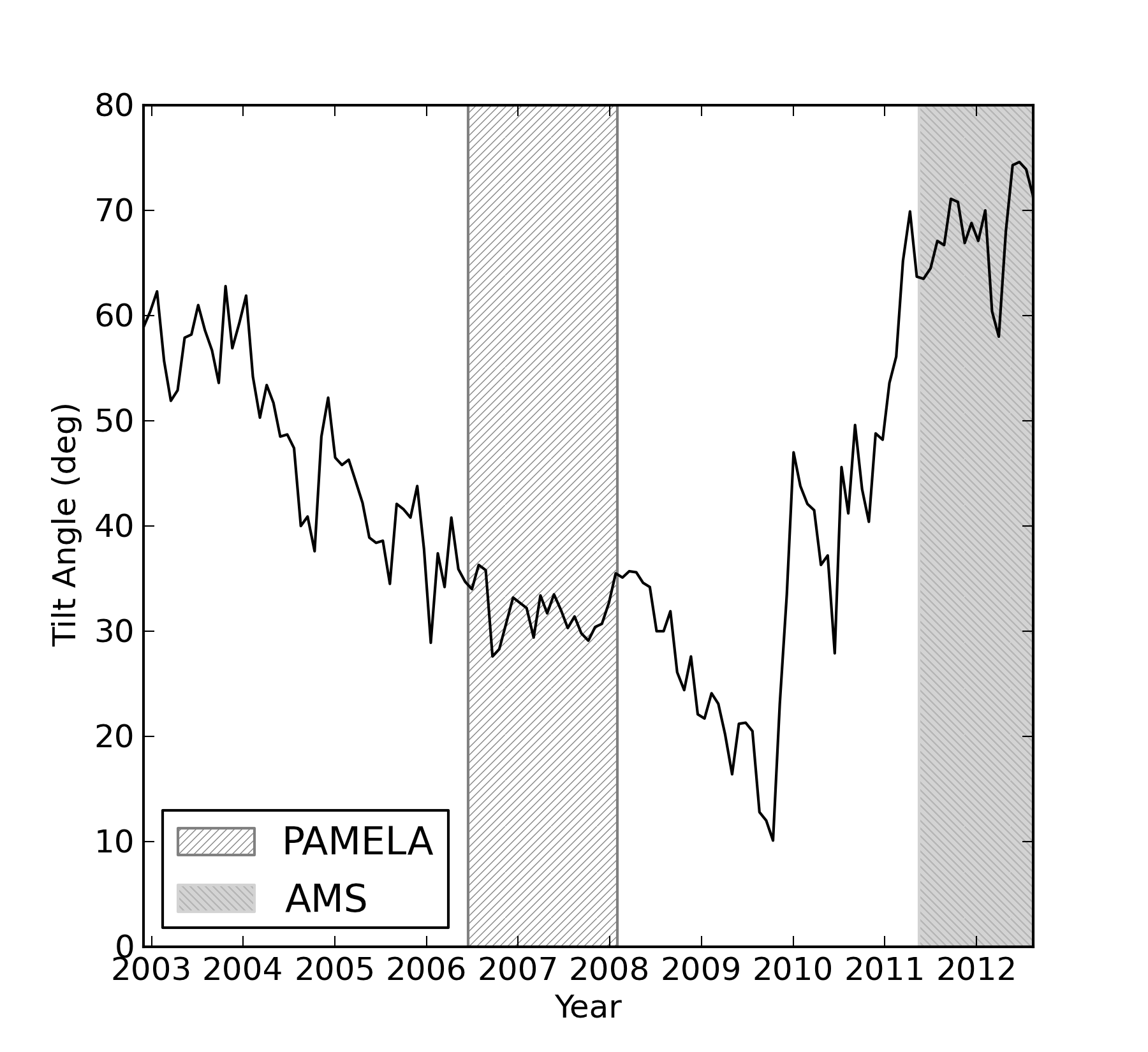}}
\end{center}
\caption{The tilt angle during data taking for PAMELA and AMS. For PAMELA we take the dates for the dataset presented in \cite{Adriani:2008zr}. Data from \cite{WSO} \label{tilts}}
\end{figure}

The tilt angle of the current sheet changes on a monthly basis. Fig.~\ref{tilts} shows the tilt angle at the sun, measured by the Wilkinson Solar Observatory \cite{WSO}. The magnetic field produced at the sun is carried to the heliopause by the solar wind. For solar wind speeds of $400-800km/s$ it takes on the order of a year for changes in the sun's magnetic field to alter the full heliospheric magnetic field. 

As well as the direct paths to the center of the heliosphere provided by the current sheet, interstellar cosmic rays can also reach the center of the solar system through a combination of diffusion, convection and drifts \cite{DiffusionReview,DiffusionReview2}. Comparison of Voyager and IMP measurements of cosmic ray fluxes show a local deficit of cosmic rays at the center of the heliosphere compared to the flux at the heliopause for low energy cosmic rays (<1GeV). There are no measurements of the suppression of the local flux in the energy range considered here (10-100GeV) and we take it to be a free parameter.

In our model, the flux of cosmic rays in the center of the heliosphere can be written:
\begin{equation}
F(E)=(\epsilon(E)+A)F_{80AU}(E).
\end{equation}
$F_{80AU}$ is the interstellar cosmic ray flux. $A$ parameterizes the local suppression of cosmic rays that reach the center of the heliosphere through convection, diffusion and drifts. $\epsilon(E)$ is the ratio of the cosmic ray flux that reaches the center of the heliosphere directly, compared to the interstellar flux. 

\section{The Positron Excess}

\begin{figure}
\begin{center}
\scalebox{0.54}{\includegraphics{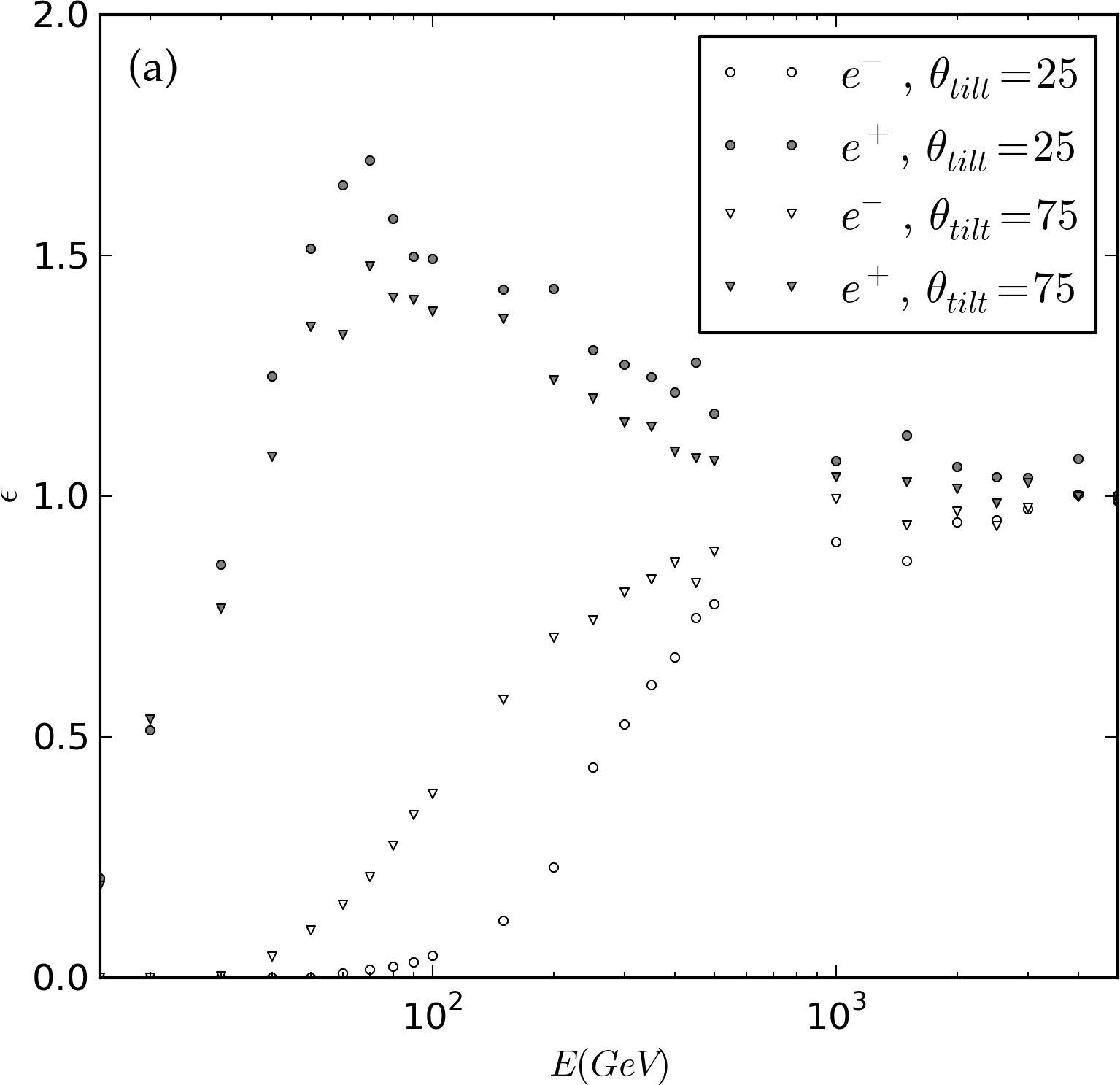}}
\scalebox{0.54}{\includegraphics{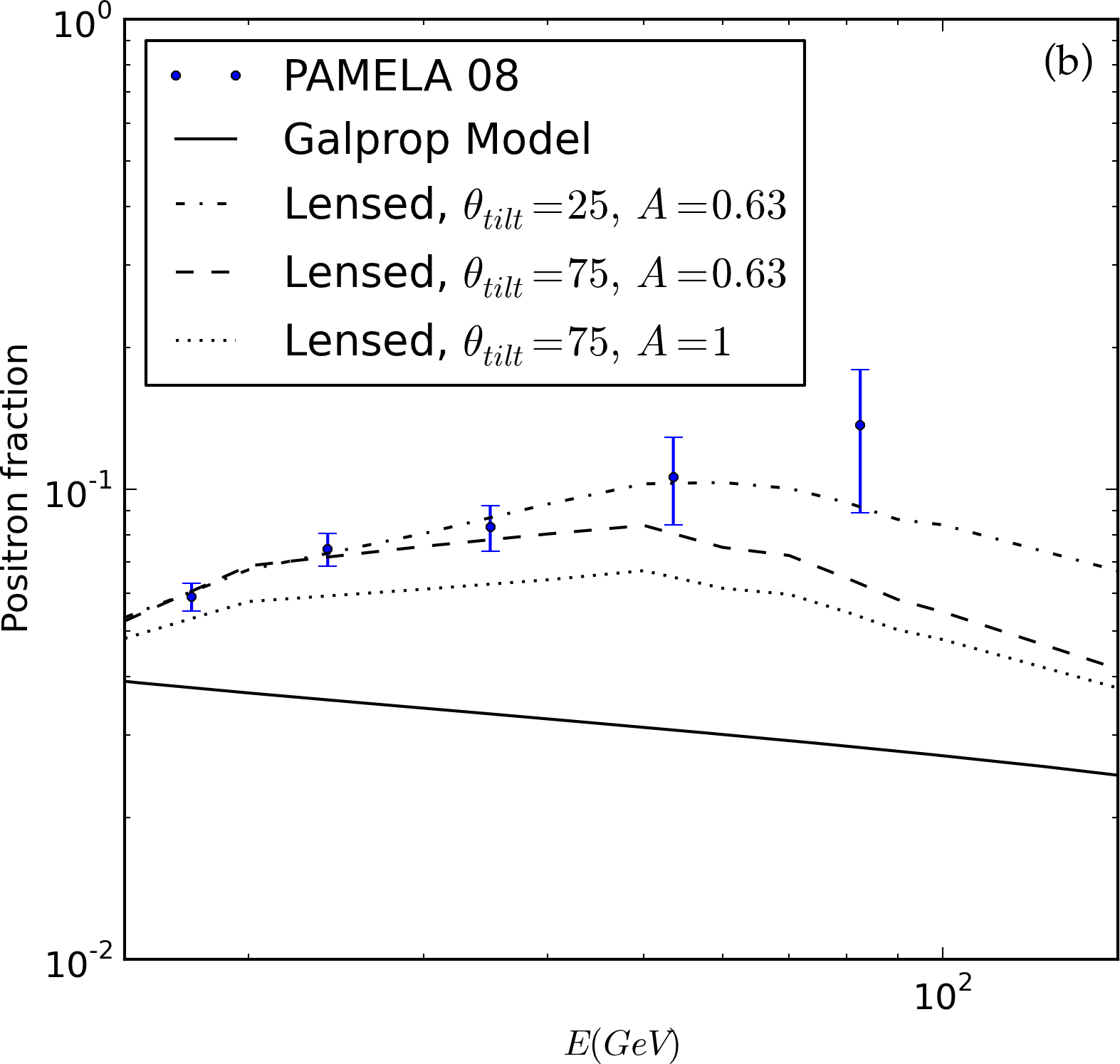}}
\end{center}
\caption{The relative flux of particles in the center of the heliosphere compared to the interstellar flux (a) and the positron fraction for different tilt angles, and local suppression factors (b). \label{fractions}}
\end{figure}

We calculate $\epsilon$ using a 2D numerical simulation of the propagation of electrons and positrons. At low energies (below a few GeV), the propagation of particles in the heliosphere must be treated as a 3D system accounting for the complex processes of convection, diffusion and drifts. At energies above a few TeV the Larmor radii of electrons and positrons are greater than the size of the heliosphere, and the magnetic field is perpendicular to their direction of travel. The paths of cosmic rays of these energies lie in a plane, and the problem can be modeled in 2D. Here we are modeling the fraction of particles in the intermediate energy range that can penetrate the heliosphere directly by crossing the current sheet.  For these particles, a 2D model can provide semi-quantitative accuracy, prior to completion of a 3D model which is under development.

We model the magnetic field as a  Parker spiral \cite{Parker} with an oscillating current sheet. In Fig.~\ref{fractions}a) we plot $\epsilon(E)$ for both a 25 degree and 75 degree tilt angle. The current sheet focuses the positrons, generating an enhancement of the positron flux in the central heliosphere that peaks at 70GeV.

We use the values of $\epsilon$ along with positron and electron spectra, generated using Galprop \cite{Galprop}, to calculate the positron fraction $F^+/(F^++F^-)$. Fig.~\ref{fractions}b) shows the resulting positron fractions. The dash-dot line shows the fit to the PAMELA data with $\theta_{tilt}=25$ where we allow $A$ to vary. The dashed line shows the positron fraction for a tilt angle of 75 degrees, with the same $A$. As $A$ may vary with time, we also show the positron fraction for $A=1$, $\theta_{tilt}=75$ (dotted line).

\section{Conclusions}

We have presented the results of a 2D numerical simulation of cosmic ray transport within the heliosphere following the model of \cite{JR PAMELA}. If the positron excess observed by PAMELA was caused by the heliospheric magnetic field, AMS should observe a smaller positron fraction than PAMELA above $40\eev$ due to the increasing tilt angle of the heliospheric current sheet.

\section*{Acknowledgments}
This work was supported by the NSF grants NSF-PHY-0701451, NSF-PHY-0900631 and NSF-PHY-0970075.

\end{document}